\documentclass[final,1p,times]{elsarticle} 

\usepackage{graphicx}
\usepackage{amssymb} 
\usepackage{amsthm} 
\usepackage{lineno}

\newcommand{\pT}{$p_{T}$}
\newcommand{\kT}{$k_{T}$}
\newcommand{\dpT}{$\delta{p}_{T}$}
\newcommand{\rhochem}{${\rho}_{ch+em}$}
\newcommand{\rhoch}{${\rho}_{ch}$}

\journal{Nuclear Physics A} 

\begin{document}

\begin{frontmatter} 

\title{Inclusive jet spectra in 2.76 TeV collisions from ALICE}

\author{Rosi Jan Reed (for the ALICE\fnref{col1} Collaboration)}
\fntext[col1] {A list of members of the ALICE Collaboration and acknowledgements can be found at the end of this issue.}
\address{Yale University, New Haven, CT 06520}


\begin{abstract} 
 Measurements of high-\pT~particle production in heavy-ion collisions at RHIC have shown that medium-induced energy loss affects the partons produced in the early stage of a heavy-ion collision.  The increased initial production cross section for partons at LHC energies makes fully reconstructed jets available in a wide kinematic range, which allows for a more differential investigation of parton energy loss.  Partonic energy loss allows us to access key observables of the hot deconfined nuclear matter produced in heavy-ion collisions.  The inclusive cross-section of reconstructed jets using the ALICE tracking detectors and electromagnetic calorimeter is presented from data collected during the 2.76 TeV pp runs. The procedures used to reconstruct jets and extract them from a fluctuating background in Pb-Pb collisions are discussed.  
  \end{abstract} 

\end{frontmatter} 


\section{Introduction}


Jets make an excellent probe of the hot, dense matter formed in heavy-ion collisions as the hard-scattered partons that fragment into jets are produced early in the collision.   The partons are modified as they propagate through the dense QCD medium and this modification is reflected in the measured jet spectrum.   It has been found that this spectrum deviates from what would be expected if the jet spectrum as measured in pp collisions is simply scaled by the number of binary collisions \cite{AtlasJetQuench}.  This deviation is called jet quenching, which is attributed to the hard scattered partons losing energy within the medium \cite{Wiedemann:2009sh}.

The results reported in these proceedings are from data collected by the ALICE experiment in 2011 after the complete installation of the ALICE Electromagnetic Calorimeter (EMCal) detector.  The pp and Pb-Pb collisions both had an energy of $2.76$ TeV per nucleon pair.  The Pb-Pb results are from the top 10\% central events where the modification due to the QCD medium is maximal.  Charged particles were reconstructed with the Inner Tracking System (ITS) and the Time Projection Chamber (TPC) and clusters were reconstructed with the EMCal.

\section{pp}

To quantify the in-medium modification of the jet spectrum, a reference measurement of the differential cross section in pp collisions at the same energy was made.  There were two different triggers in the data used to determine the pp cross-section.  One trigger was the minimum bias (MB) trigger, which requires at least one hit in either of the VZEROS, highly forward scintillator detectors, or in the Silicon Pixel Detector (SPD) which is at mid-rapidity.  The second was from EMCal triggered events, which required the MB trigger condition and an additional single shower in the EMCal with ${E}_{T}>$ 3 GeV.  The shower signal is a sum of energy in groups of 4x4 neighboring EMCal towers.  This extended the \pT~reach of the pp analysis.

A clustering algorithm that combined signals from 3$\times$3 neighboring EMCal towers determined the EMCal clusters.  To correct for energy deposits of charged particles in the EMCal the tracks were extrapolated to the EMCal and matched to clusters in the window $|\eta| < $0.015 and $|\phi| <$ 0.03 and up to 100\% of the matched track momentum is subtracted from the cluster. The cluster energies after hadronic correction and the tracks are used as inputs to the jet finder for both the pp and Pb-Pb results.

For these analyses the jet collection was determined by the anti-\kT~algorithm \cite{JetFinder1} with resolution parameters $R$ = 0.2 and $R$ = 0.4.  The minimum \pT~of the track jet constituents is 0.15 GeV/c and the minimum ${E}_{T}$ for the clusters is 0.15 GeV.  In heavy-ion collisions it is necessary to determine the contribution to the jet from the underlying event.  This is done by calculating the underlying event momentum density, \rhochem~from the collection of tracks and EMCal clusters. For the Pb-Pb analysis, \rhochem~was calculated from background jets resulting from the \kT~algorithm \cite{JetFinder2} run with resolution parameter $R$ = 0.2.  In all cases, only those anti-\kT~or \kT~jets that were fully contained within the EMCal acceptance were used.  This was defined by requiring the jet axis to be a distance $R$ away from the EMCal boundaries of $|\eta| <$ 0.7 and 1.4 $<\phi < \pi$. 

For the jet cross-section analysis, the jets were corrected back to the particle level utilizing a bin-by-bin technique, which is simulation based.  This was determined by comparing the jet spectra from the particle level to the spectra that results when detector corrections are applied by use of GEANT.  The contributions to the correction can be placed in two categories, effects that smeared the jet energy or effects that shifted the jet energy.  The main sources that contribute to the jet energy shift are unmeasured neutral particles, the tracking efficiency and the hadronic correction.  The jet energy smearing results from the detector resolution and event-by-event fluctuations of the jet energy shift.

\begin{figure}[htbp]
\begin{center}
 \includegraphics[width=0.8\textwidth]{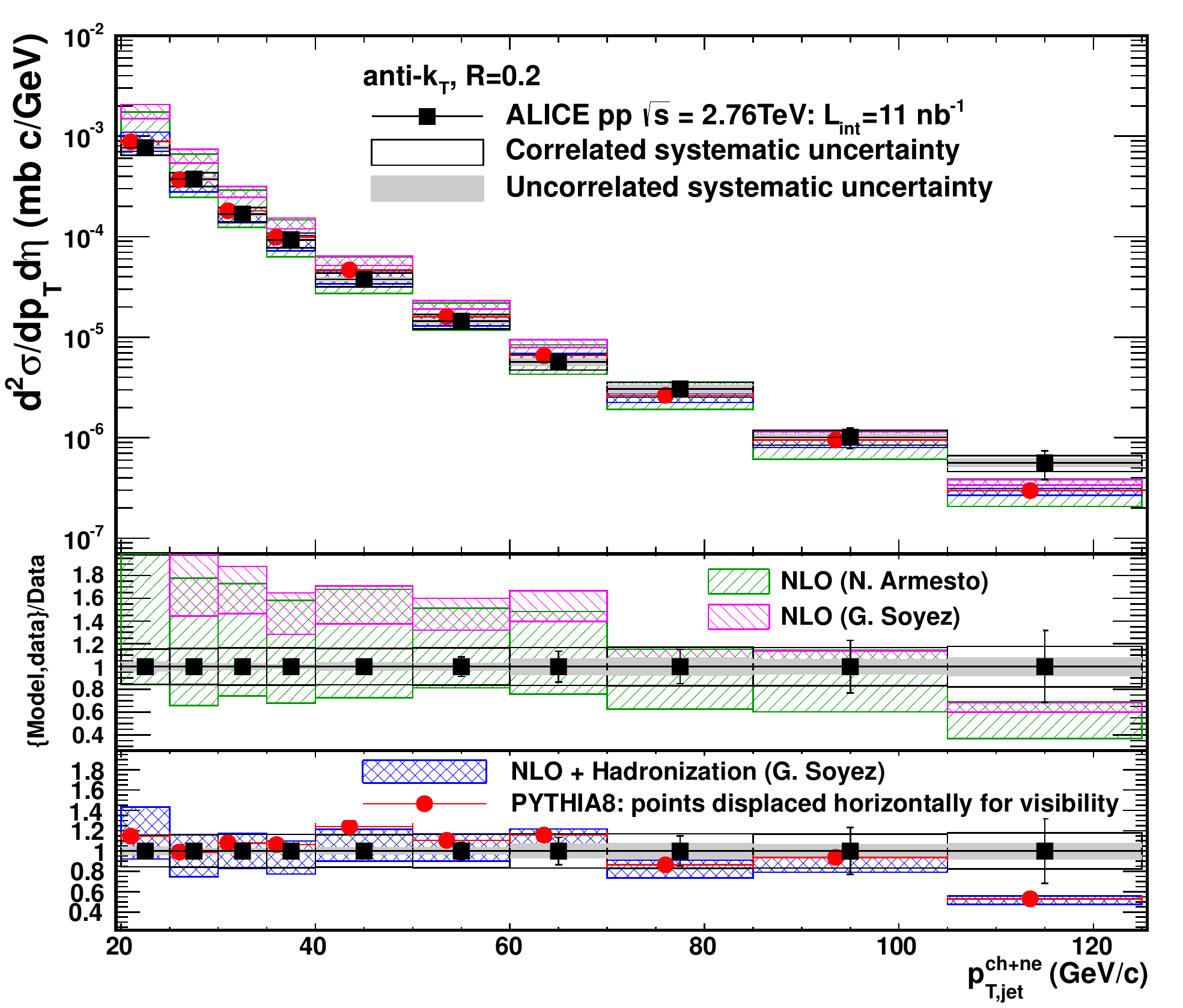}
 \end{center}
\caption{The solid boxes indicate the full inclusive jet differential cross section obtained with resolution parameter $R$ = 0.2 measured in pp collisions at 2.76 TeV data \cite{RMa}.  The solid circles represent a PYTHIA8 prediction.  The solid boxes are the uncorrelated systematic uncertainties on the cross-section, the open boxes are the correlated systematic uncertainties.  The hashed boxes indicate the NLO calculations (color online).}
\label{fig:ppPlot}
\end{figure}

Figure \ref{fig:ppPlot} shows the inclusive differential jet cross section obtained with a resolution parameter of $R$ = 0.2. Also shown are the results of a pQCD calculation at NLO and a PYTHIA8 prediction.  The uncertainties for the NLO calculations were determined by adding the renormalization and factorization scale uncertainties, estimated by varying the scales from 0.5 to 2 ${p}_{T}$, in quadrature \cite{RMa}.  

\section{Pb-Pb}

One of the main experimental challenges of measuring the jet spectrum in heavy-ion collisions is removing the contribution from the underlying event.  In order to fully quantify the background contribution from the underlying event we need to determine the average background energy density, \rhochem, and the width of the point-to-point background fluctuations, characterized as \dpT.   \rhochem~was calculated as the median of ${p}_{T,jet}/{A}_{jet}$, where ${A}_{jet}$ is the area of the jet, of the \kT~jets found within the EMCal.  This background was subtracted from the reconstructed momentum of the anti-\kT~signal jets using the formula ${p}_{T,jet} = {p}_{T}^{rec} - \rho\cdot{A}_{jet}$ \cite{JetFinder3, JetFinder4}. 

The effect of point-to-point fluctuations in the background on the measured jet spectrum is quantified through the calculation of \dpT.   This can be determined e.g. by placing random cones in the measured events or by embedding a high \pT~probe in the event and then looking at the collection of jets found by the anti-\kT~algorithm that contain the embedded probe.  The distribution of \dpT~for a particular centrality is then calculated as \dpT~$= {p}_{T}^{rec} - \rho\cdot{A}_{jet} � {p}_{T}^{probe}$ \cite{ALICEbackground}.  The width of the distribution was determined to be 6.1 GeV/$c$ for both techniques, compared to 4.5 GeV/$c$ measured with a charged track only analysis.

\begin{figure}[htbp]
\begin{center}
\mbox{ \includegraphics[width=0.45\textwidth, height = 0.25\textheight]{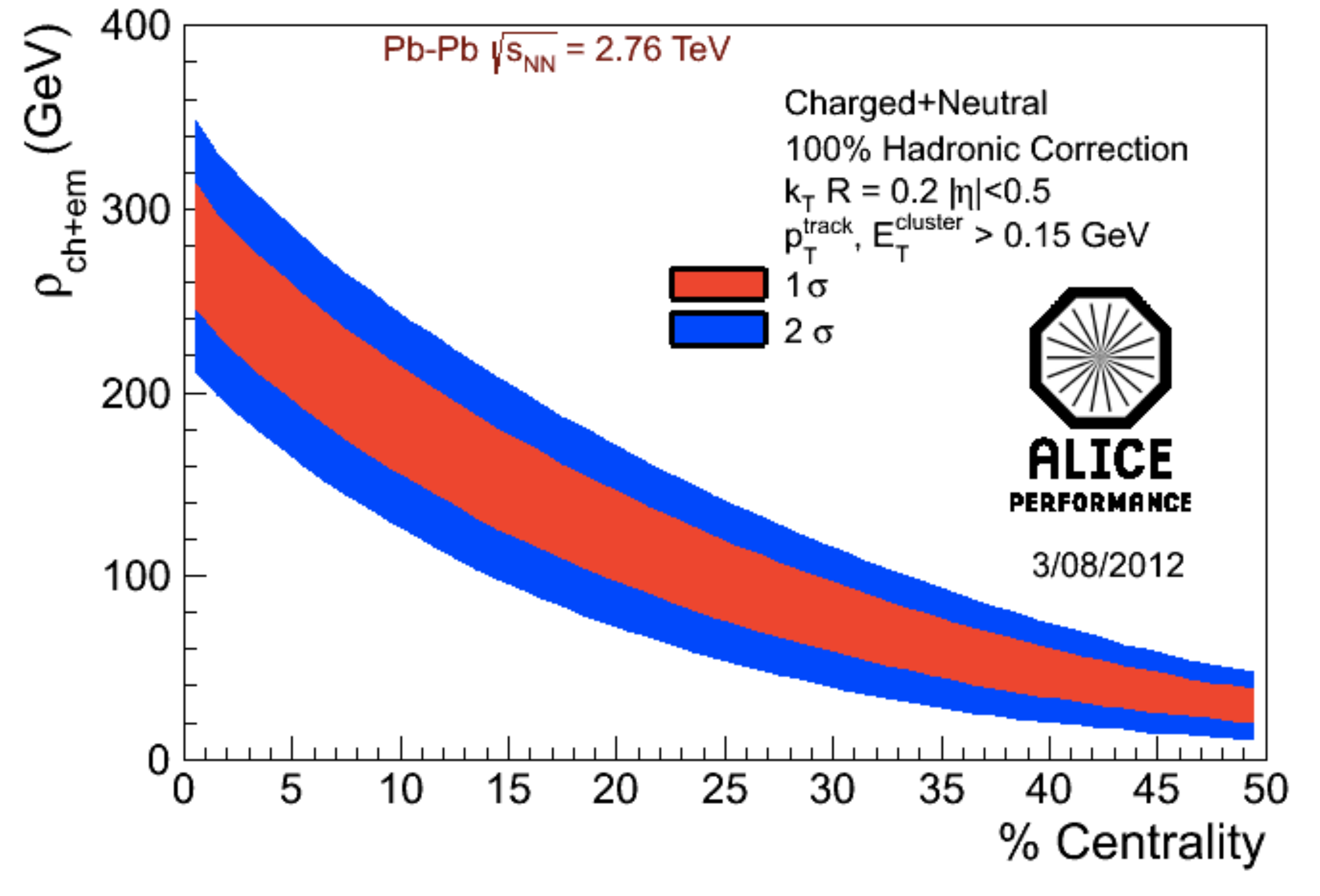}}
\mbox{ \includegraphics[width=0.45\textwidth, height = 0.25\textheight]{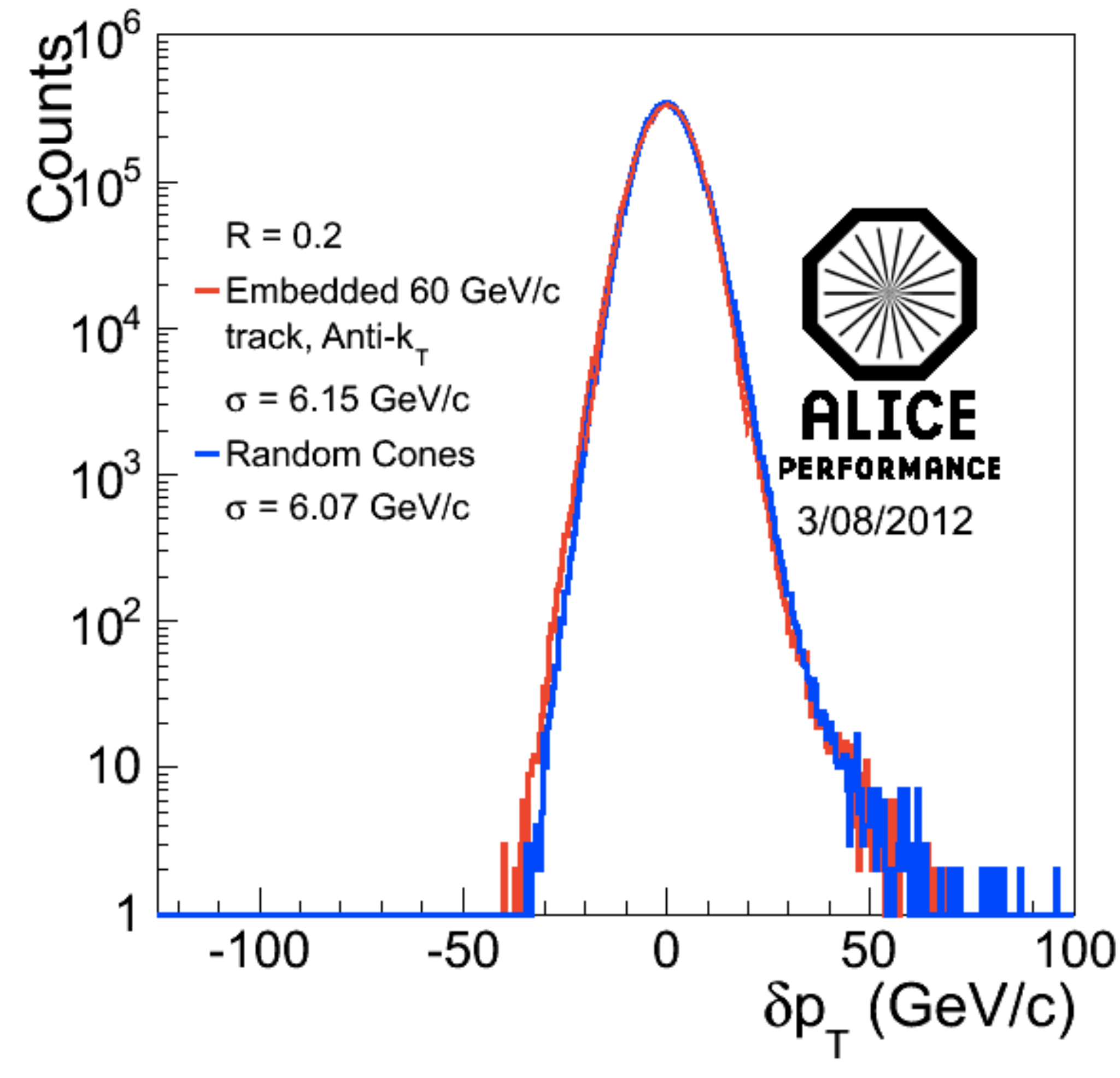}}
 \end{center}
\caption{On the right is the distribution of \rhochem~versus centrality calculated with \kT~jets determined with a resolution parameter of 0.2.  The inner colored band indicates the 1$\sigma$ distribution and the outer colored band indicates the 2$\sigma$ distribution. Shown on the left is the \dpT~distribution for two different methods with characteristic size $R$ = 0.2.  The red distribution is the result from embedding a high \pT~object and determining \dpT~of the anti-\kT~jets that contain the embedded object.  The blue distribution is determined by randomly throwing random cones within the EMCal acceptance with $R$ = 0.2.}
\label{fig:rhochem}
\end{figure}

In Figure \ref{fig:rhochem} we show the \rhochem~distribution determined by taking the median of \kT~jets found with a resolution parameter of 0.2, which contain both tracks and clusters in the 50\% most central events.  The 1$\sigma$ and 2$\sigma$ widths are also indicated.   This shows the event-to-event variation in the average energy density within the EMCal acceptance for a given centrality.  As expected \rhochem~is larger than the value of \rhoch~quoted in \cite{ALICEbackground} due to the inclusion of the neutral energy component through the addition of the calorimeter clusters.  Figure \ref{fig:rhochem} also shows the \dpT~determined from the two different methods outlined above.  The random cones had a radius of 0.2 which corresponds to the average size of a jet.  The embedded object was a 60 GeV/c cluster and the jet finder used to determine this distribution was the anti-\kT~algorithm run with a resolution parameter of 0.2.  Both of these algorithms yield very similar results, and indicate that the \dpT~increased by roughly 50\% from \dpT~calculated by charged tracks only \cite{ALICEbackground}.  This increase in the average energy and the background fluctuations due to the addition of the calorimeter clusters increases the difficulty of analyzing fully reconstructed jets.  Results from an analysis of charged jets can be found in the proceedings at reference \cite{MartaHP}.

\section{Outlook}
In this proceedings we outlined how the background characterization methods developed for jet analyses using only charged jets can be applied to full jet analyses with the addition of calorimeter clusters \cite{ALICEbackground}, which is one of the main experimental challenges to measuring fully reconstructed jets in heavy-ion events.  The techniques discussed in reference \cite{MartaHP} will be applied and we will measure a fully reconstructed jet spectrum in the Pb-Pb data set described in this proceedings.  Together with the reference pp measurement reported here we will then be able to report a \pT~dependent ${R}_{AA}$.

\end{document}